\documentclass[a4paper,fleqn,usenatbib,useAMS]{mnras}
\usepackage{graphicx}	% Including figure files
\usepackage{amsmath}	% Advanced maths commands
\usepackage{amssymb}	% Extra maths symbols
\usepackage{multicol}        % Multi-column entries in tables
\usepackage{bm}		% Bold maths symbols, including upright Greek
\usepackage{pdflscape}	% Landscape pages
\usepackage{times}
\usepackage{threeparttable}
\usepackage{color}
\usepackage{ulem}

\begin{document}

% write title (with email and institute)
\title{A Blind Search for Prompt Gamma-ray Counterparts of 
Fast Radio Bursts with \textit{Fermi}-LAT Data}
\author[Yamasaki, Totani, \& Kawanaka]
{Shotaro Yamasaki\thanks{E-mail: yamasaki@astron.s.u-tokyo.ac.jp}, Tomonori Totani, and Norita Kawanaka \\
Department of Astronomy, School of Science, The University of Tokyo, Hongo, Bunkyo-ku, Tokyo 113-0033}
\maketitle

\begin{abstract}
Fast Radio Bursts (FRBs) are a mysterious flash phenomenon detected in
radio wavelengths with a duration of only a few milliseconds, and they
may also have prompt gamma-ray flashes. Here we carry out a blind
search for msec-duration gamma-ray flashes using the 7-year
\textit{Fermi} Large Area Telescope (\textit{Fermi}-LAT) all-sky
gamma-ray data. About 100 flash candidates are detected, but after
removing those associated with bright steady point sources, we find no
flash events at high Galactic latitude region ($|b|>20^{\circ}$).
Events at lower latitude regions are consistent with statistical
flukes originating from the diffuse gamma-ray background.  From these
results, we place an upper limit on the GeV gamma-ray to radio flux
ratio of FRBs as $\xi \equiv \left(\nu L_{\nu})_{\gamma}\right/ (\nu
L_{\nu})_{\rm radio} \lesssim (4.2$--$12)\times10^7$, depending on the
assumed FRB rate evolution index $\beta=0$--$4$ [cosmic FRB rate
  $\Phi_{\rm{FRB}}\propto (1+z)^{\beta}$]. This limit is comparable
with the largest value found for pulsars, though $\xi$ of pulsars is
distributed in a wide range.  We also compare this limit with the
spectral energy distribution of the 2004 giant flare of the magnetar
SGR 1806$-$20.

\end{abstract}

% look up list of allowable keywords for this section

\section{Introduction}
Fast radio bursts (FRBs) are a population of radio transients with
durations of only about 1 msec.  Since the discovery of the so-called
Lorimer Burst \citep{2007Sci...318..777L}, 17 FRBs have been reported
by radio transient surveys to date (15 by Parkes radio telescope,
\citealt{2007Sci...318..777L}; \citealt{2012MNRAS.425L..71K};
\citealt{2013Sci...341...53T}; \citealt{2014ApJ...792...19B};
\citealt{2015MNRAS.454..457P}; \citealt{2015ApJ...799L...5R};
\citealt{2015arXiv151107746C}; \citealt{Keane:2016fk}; 1 by
Arecibo, \citealt{2014ApJ...790..101S}; \citealt{Spitler:2016uq}; \citealt{2016arXiv160308880S}; and 1 by Green Bank Telescope,
\citealt{Masui:2015fk}).  Their dispersion measures (DMs) are much
larger than those expected for objects in our Galaxy, and a
cosmological distance scale of $z \sim$ 0.5--1 is inferred if the
dominant contribution to DMs is from electrons in intergalactic
meduim.  A recent study of FRB150418 has identified its radio
afterglow associated with an elliptical galaxy, whose redshift ($z =
0.492$) is consistent with that expected by its DM
\citep{Keane:2016fk}.  However, no counterparts in other
wavelengths have been detected.  There are various progenitor models
proposed for FRBs; some of them are related to young stellar
populations, while others to old.  The former includes giant flares
from soft gamma-ray repeaters (SGRs;
\citealt{popov2010hyperflares1465953,2013Sci...341...53T,2014MNRAS.442L...9L,2014ApJ...797...70K}),
collapsing supermassive neutron stars (``blitzar'' model;
\citealt{2014A&A...562A.137F}), and giant radio pulses from pulsars
\citep{2015arXiv150505535C,2015arXiv150100753C}.  The latter includes
binary neutron star (or black hole) mergers
\citep{2013PASJ...65L..12T,2015ApJ...814L..20M}, and binary white
dwarf mergers \citep{2013ApJ...776L..39K}.

Whatever the progentor is, it is reasonable to expect corresponding
msec-duration radiations of FRBs in other wavelengths. Here we search
for gamma-ray counterparts of FRBs using the 7-year \textit{Fermi}
Large Area Telescope (\textit{Fermi}-LAT) data.  Radio telescopes
finding FRBs at cosmological distances have narrower field-of-views
($\sim 15$ arcmin beam size for Parkes) than \textit{Fermi}-LAT, and
hence we expect that \textit{Fermi}-LAT detects FRBs at smaller
distances than those found in radio bands.  Therefore, rather than
examining the \textit{Fermi}-LAT data at known FRB locations, we
perform a blind search of msec-duration gamma-ray flashes (MGFs),
without prior information about FRBs in radio bands.  Then we can
derive more stringent constraints on the GeV gamma-ray to radio flux
ratio of FRBs, than that from upper limits on gamma-ray flux set by
{\it Fermi}-LAT for individual radio-detected FRBs.

The outline of this paper is as follows.  In section \ref{sec:Data and
  Search Method}, we describe the details of the gamma-ray data set
and our search method.  We then present the resuls of the MGF search
in section \ref{sec.Result}. Constraints on the gamma-ray to radio
flux ratio of FRBs will be derived in section \ref{sec:Implications
  for FRBs}, followed by discussion for implications.  Throughout this
paper, we assume a flat-universe $\Lambda$CDM model with
$H_0=67.8\rm{\:\:km\:s^{-1}\:Mpc^{-1}}$, $ \Omega_M=0.308$,
$\Omega_{\Lambda}=0.692$ \citep{2015arXiv150201589P}

\section{Data and Search Method}
\label{sec:Data and Search Method}

The {\it Fermi}-LAT is a pair-conversion gamma-ray telescope designed
to cover the energy band from 20 MeV to greater than 300 GeV.  In this
work we use the Pass 7 Reprocessed weekly data publicly available at
the Fermi mission
website\footnote{http://fermi.gsfc.nasa.gov/ssc/data/access/}.  The
analysis spans the time period of 6.8 years from July 31, 2008 (UTC)
to June 18, 2015 (UTC), corresponding to the mission elapsed time
(MET) from 239557417 s to 455059763 s and the Fermi mission week from
1 to 368. We use P7SOURCE class photons with reconstructed energies
from 1 to 100 GeV.  We used the Fermi Science Tools version v9r32p5
package\footnote{http://fermi.gsfc.nasa.gov/ssc/data/analysis/software/v9r32p5.html}
and the P7REP instrument response functions (IRFs) in our analysis.
In accordance with the data analysis procedure outlined in Fermi
mission website, we firstly selected 1 GeV$<E<$100 GeV photons of
``P7SOURCE'' class.  There is another ``Transient class'', which is
for gamma-ray events with looser cuts than P7SOURCE and hence includes
more background particle contamination.  Though this class is useful
for bright transients during which background contamination is
negligible, we use P7SOURCE because we will search faint (i.e., small
number statistics) transient events, to which background contamination
is crucial.  We also set the Earth relative zenith angle cut of
$100^{\rm{\circ}}$ to reduce the bright earth-limb gamma-rays.
Finally we select gamma-rays only in the good time intervals (GTIs)
when the data quality is good, by excluding the bad time intervals
(BTIs) due to the spacecraft events. The total amount of GTIs is 79 \%
of the MET.

In order to search for MGF candidates, we apply a search algorithm as
follows.  First we define the search time window $\Delta t$, and four
values of $\Delta t = $ 1, 2, 5, and 10 msec are tried here.  We
consider a reference gamma-ray event, and other gamma-ray events are
searched within $2^{\circ}$ radius from the reference in the time
interval of $\Delta t$ starting from the reference event. The radius is
determined considering the per-photon angular resolution, $\sim
0.8^{\circ}$ (68\% containment of the point-spread function) at 1 GeV
\citep{2009ApJ...697.1071A,2015ApJS..218...23A}.  The number of
gamma-rays in this time window and the circle, including the first
reference photon itself, is denoted as $N_{\rm{ph}}$.  Then we repeat
this procedure for all the gamma-ray events of the Fermi data set that
we use.  It is possible that a gamma-ray event is included in
different MGFs as defined above, but we present all MGF candidates as
different events in the following if they are different as a
combination of photons.  The detector deadtime per event of {\it
  Fermi}-LAT is $<100\,\rm{\mu s}$, and it is negligible in our
analysis.

\section{Result}
\label{sec.Result}
\subsection{MGFs Associated with Known Gamma-ray Sources}
\label{subsec:PS MGFs}

The result of our blind search for MGFs is shown in Table
\ref{table:MGFs} (see columns of ``No cut").  For $N_{\rm ph} = 2$,
there are 17, 33, 68 and 133 MGF candidates for $\Delta t = 1$, 2, 5
and 10 ms, respectively, as shown in the Table.  Only one MGF event
was found for $N_{\rm ph} = 3$ and $\Delta t = $ 5 or 10 ms, and there
were no events of $N_{\rm{ph}}\ge4$ for any value of $\Delta t$.  We
examined the $N_{\rm ph} = 3$ event, and found that it is associated
with GRB 090510 \citep{2010ApJ...716.1178A}, which shows photon
clusterings within a time-scale of $< 10$ ms.  The total duration of
detected LAT gamma-rays for GRB 090510 is 0.1s, which is much longer
than FRBs. Therefore we remove this event from our sample in the
following, and the results after this ``1st cut'' are shown in Table
\ref{table:MGFs}. Only $N_{\rm ph} = 2$ events remain.

In order to examine the possibility that the detected $N_{\rm{ph}}=2$
events are caused by bright point sources, we examined the nearest
point source from each MGF event.  We use the Fermi Large Area
Telescope Third Source Catalog (3FGL; \citealt{2015ApJS..218...23A})
which consists of $\sim3000$ gamma-ray sources.  Figure \ref{Fig:
  catalog flux} presents the photon flux distribution of the nearest
point sources, in comparison with that of all the 3FGL catalog point
sources.  The 3FGL sources nearest to the MGF candidates have
obviously much brighter fluxes compared with the general 3FGL sources,
indicating that most MGF candidates are caused by bright point
sources.  Figure \ref{fig: catalog distance} is the histogram of
angular distance from the MGF candidates to the nearest 3FGL sources.
The angular distribution is similar to the \textit{Fermi}-LAT point
spread functions (PSF;
\citealt{2009ApJ...697.1071A,2015ApJS..218...23A}), and much smaller
than the distribution expected for the case that the MGF candidates
are randomly distributed in the sky, again indicating that these are
caused by bright point sources.  The classes of the nearest sources
include a variety of populations for which msec scale variability is
physically unlikely (e.g.  blazars), and the most likely origin of the
MGF candidates is simply statistical flukes induced by the Poisson
statistics of a steady gamma-ray flux.  Indeed, we confirmed that the
number expected by Poisson statistics and point source fluxes is in
rough agreement with that of the detected MGF candidates.

There is an excess of the MGF distribution in Fig.  \ref{fig: catalog
  distance} at $\theta \gtrsim 2.0^{\circ}$ compared with PSF, which
implies a contribution of MGFs that are not caused by point sources.
We will see that these are likely caused by the diffuse gamma-ray
background (see below).  To remove MGF candidates related to bright
point sources, MGF candidates are excluded from the final sample if
their nearest 3FGL source flux is larger than $4\times10^{-9}
\:\rm{photons/cm^{2}/s}$ and angular separation is smaller than
$2.0^{\circ}$.  The number of 3FGL sources exceeding this flux
threshold is 293.  The result of this cut is shown in Table
\ref{table:MGFs} as ``2nd cut''.

\begin{table*}
	\smallskip 
	\begin{center}
	\caption{The number of detected MGF candidates
          ($N_{\rm{obs}}^{\rm{2ph}}$ and $N_{\rm{obs}}^{\rm{3ph}}$ for
          $N_{\rm ph} = 2$ and 3, respectively).  ``No cut" is the
          result of our blind search for MGFs without any event cut.
          The ``1st cut'' is to remove MGF candidates associated with
          GRB 090519.  The ``2nd cut'' is the result after removing
          MGF events that are caused by nearby bright sources
          [brighter than a photon flux of $4 \times 10^{-9}
            \,\rm{photons\;cm^{-2}\;s^{-1}}$ (1 GeV $<E<$ 100 GeV)
            with angular separation less than 2.0$^\circ$].  The ``3rd
          cut'' is removing regions close to the Galactic disk ($|b| <
          20^\circ$).  The expected numbers of false events caused by
          the Poisson statistics of the diffuse gamma-ray background
          flux are shown as $N_{\rm pred}^{\rm 2ph}$ in the 2nd and
          3rd cut results.}
	\label{table:MGFs}
	\begin{threeparttable}
	\begin{tabular}{clccccccccccc}
\hline\hline
Cuts && \multicolumn{2}{c}{No cut}& & \multicolumn{2}{c}{1st cut}& & \multicolumn{2}{c}{2nd cut}& & \multicolumn{2}{c}{3rd cut}\\
\cline{3-4} \cline{6-7}  \cline{9-10}  \cline{12-13}
$f_{\rm{sky}}$\tnote{a}&& \multicolumn{2}{c}{1.0}& & \multicolumn{2}{c}{1.0}& & \multicolumn{2}{c}{0.96}& & \multicolumn{2}{c}{0.68}\\
\cline{3-4} \cline{6-7}  \cline{9-10}  \cline{12-13}\\
&&$N_{\rm{obs}}^{\rm{2ph}}	$&$N_{\rm{obs}}^{\rm{3ph}}$&&$N_{\rm{obs}}^{\rm{2ph}}$&$N_{\rm{obs}}^{\rm{3ph}}$&&$N_{\rm{obs}}^{\rm{2ph}}\;(N_{\rm{pred}}^{\rm{2ph}})$&$N_{\rm{obs}}^{\rm{3ph}}$&&$N_{\rm{obs}}^{\rm{2ph}}\;(N_{\rm{pred}}^{\rm{2ph}})$&$N_{\rm{obs}}^{\rm{3ph}}$\\
\hline 
$\Delta t =1\rm{\:\:\:ms}$&	&17		&0  		&&14	&0 			&&1$\;$(2.7)  	&0		&&0$\;$($1.2\times10^{-1}$)  	&0\\
$\Delta t =2\rm{\:\:\:ms}$&	&33		&0 		&&29	&0			&&4$\;$(5.3)	&0  		&&0$\;$($2.3\times10^{-1}$)   	&0\\
$\Delta t =5\rm{\:\:\:ms}$&	&68		&1  		&&62 	&0			&&17$\;$(13) 	&0		&&0$\;$($5.8\times10^{-1}$)   	&0\\
$\Delta t =10 \rm{\:ms}$&	&133	&1	 	&&127	&0 			&&38$\;$(27)	&0		&&0$\;$($1.0$)			 	&0\\
\hline
\end{tabular}
		\begin{tablenotes} 
 \item[a] The remaining sky fraction after each event cut.
   		\end{tablenotes}
  	\end{threeparttable}
	\end{center}
\end{table*}

\begin{figure}
 \includegraphics[width=\columnwidth]{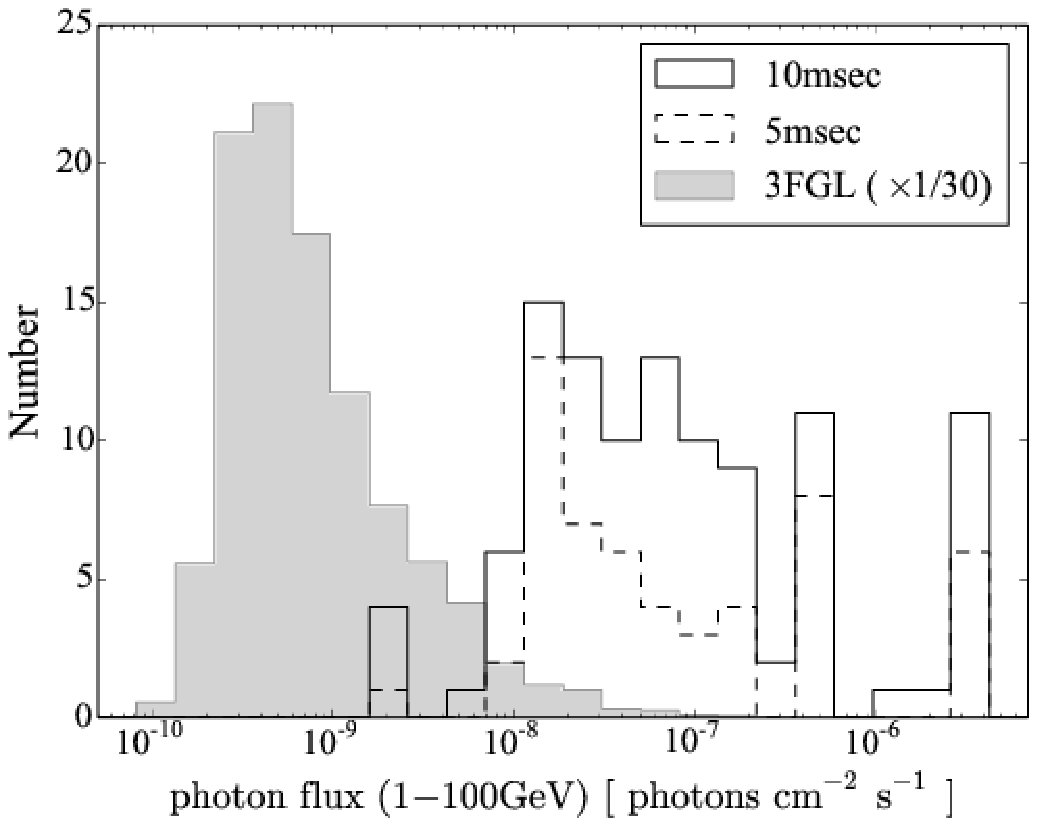}
 \caption{The photon flux (1 GeV $<E<$ 100 GeV) distribution of the
   nearest point source to each MGF candidate with $N_{\rm ph} = 2$
   and $\Delta t$ = 5 or 10 msec, in comparison with that of all
   sources in the 3FGL catalog. The histogram for the 3FGL sources is
   multiplied by a factor of 1/30 for the presentation purpose.  }
 \label{Fig: catalog flux}
\end{figure}

\begin{figure}
 \includegraphics[width=\columnwidth]{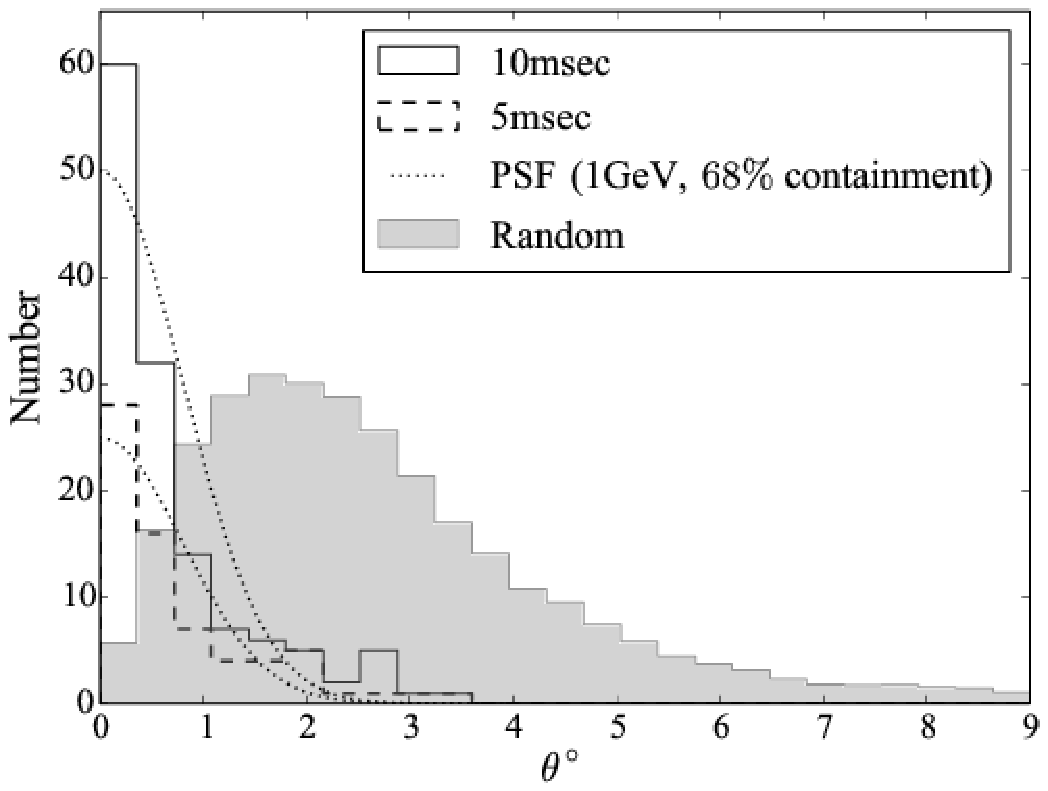}
 \caption{Angular separation distribution from each MGF candidate
   ($N_{\rm ph}=2$) to the nearest 3FGL catalog source, for $\Delta t
   =5$ and 10 ms.  The \textit{Fermi}-LAT point spread function
   profile (at 1 GeV, 68\% containment) is also shown by the dashed
   curves (two same profiles scaled to the histograms of two $\Delta
   t$ values).  For comparison, the expected distribution when MGFs
   are randomly distributed on the sky is also shown.  }
 \label{fig: catalog distance}
\end{figure}

\subsection{False MGFs by the Diffuse Background}
\label{subsec:BG MGFs}

Even after removing the regions affected by bright point sources, MGF
candidates remain.  These may be statistical flukes caused by Poisson
statistics of constant gamma-ray flux of the diffuse gamma-ray
background.  Here we estimate the number of MGF candidates expected by
this.  For this purpose, we firstly divide all sky into 72$\times$36
regular square pixels in x- and y-axis directions on the Aitoff
projection (i.e., a pixel scale of $\sim 5^\circ$), and the spatial
direction and surface area of the $j$-th pixel are denoted by solid
angle $\Omega_j$ and $\Delta \Omega_j$.  We then create a count map
and an exposure map of gamma-rays selected by the criteria described
in Section \ref{sec:Data and Search Method} using the {\it Fermi}
analysis tool of {\it gtltcube} and {\it gtexpcube}.  By dividing the
count map (in photons) by the exposure map (in $\rm{cm}^{2}\:\rm{s}$)
and the surface area $\Delta \Omega_j$, we get an intensity map (in
photons $\rm{cm}^{-2}\:\rm{s}^{-1} \ \rm{sr}^{-1}$) of the diffuse
gamma-ray background, $J(\Omega_j)$.

Next, we divide the whole mission time into time bins of $\Delta T
=10$ min, and the $i$-th time bin is denoted as $T_i$.  The exposure
map $\varepsilon(T_i, \Omega_j)$ for each time bin is calculated with
the same angular pixel binning on the sky.  We need to calculate the
effective area of the detector in each time bin to estimate the
expected number of Poisson-induced MGF events.  {\it Fermi} typically
surveys 2$\pi$ sr in 1.5 hr, and change of effective area for a given
pixel in 10 min is not large and hence we approximate it to be
constant in each time bin.  It should be noted that the effective area
is relatively flat in the gamma-ray energy range adopted in this work,
and we assume that it is constant against gamma-ray energy.  We
calculate the exposure time $T_{\rm exp}(T_i)$ in each time bin by
adding all the GTIs.  Then we can estimate $A_{\rm{eff}}(T_i,
\Omega_j)$ from exposure $\varepsilon(T_i, \Omega_j)$ calculated by
the {\it Fermi} Science Tool, from the relation
\begin{eqnarray}
\varepsilon(T_i, \Omega_j) = A_{\rm{eff}}(T_i,
\Omega_j)\:T_{\rm{exp}}(T_i),
\end{eqnarray}
where $A_{\rm{eff}}$ is the effective area averaged about energy and
angular resolutions.

Then the expected mean photon count within the MGF search time window
$\Delta t$ and the search radius (2 deg) is
\begin{eqnarray}
\lambda(T_i, \Omega_j) = 
J(\Omega_j) A_{\rm{eff}}(T_i, \Omega_j) \; \omega \; \Delta t \ ,
\end{eqnarray}
where $\omega$ is the solid angle of the search circle.  In the
Poisson statistics, the probability that $N_{\rm{ph}}$ photons
arrive is
\begin{eqnarray}
P(T_i, \Omega_j; N_{\rm{ph}}, \Delta t) = \frac{\lambda(T_i,
  \Omega_j)^{N_{\rm{ph}}}}{N_{\rm{ph}}!}  \exp[- \lambda(T_i,
  \Omega_j)].
\end{eqnarray}
The expected number of MGFs in a time bin $T_i$ at a pixel $\Omega_j$
is then obtained by multiplying the number of trials, i.e., the number
of time windows, $T_{\rm{exp}}(T_i)/\Delta t$, and the number of
search circles, $\Delta \Omega_j/\omega$:
\begin{eqnarray}
N(T_i, \Omega_j; N_{\rm{ph}}, \Delta t) =
P(T_i, \Omega_j; N_{\rm{ph}}, \Delta t)
\frac{T_{\rm{exp}}(T_i)}{\Delta t}
\frac{\Delta \Omega_j}{\omega}.
\end{eqnarray}
Then the total expected number of Poisson-induced MGF events can be
calculated by summing up $N(T_i, \Omega_j; N_{\rm{ph}}, \Delta t)$ for
$T_i$ and $\Omega_j$. Note that the event cuts introduced in the
previous section can be taken into account by removing regions around
bright point sources from $\Delta \Omega_j$.

\begin{figure}
 \includegraphics[width=\columnwidth]{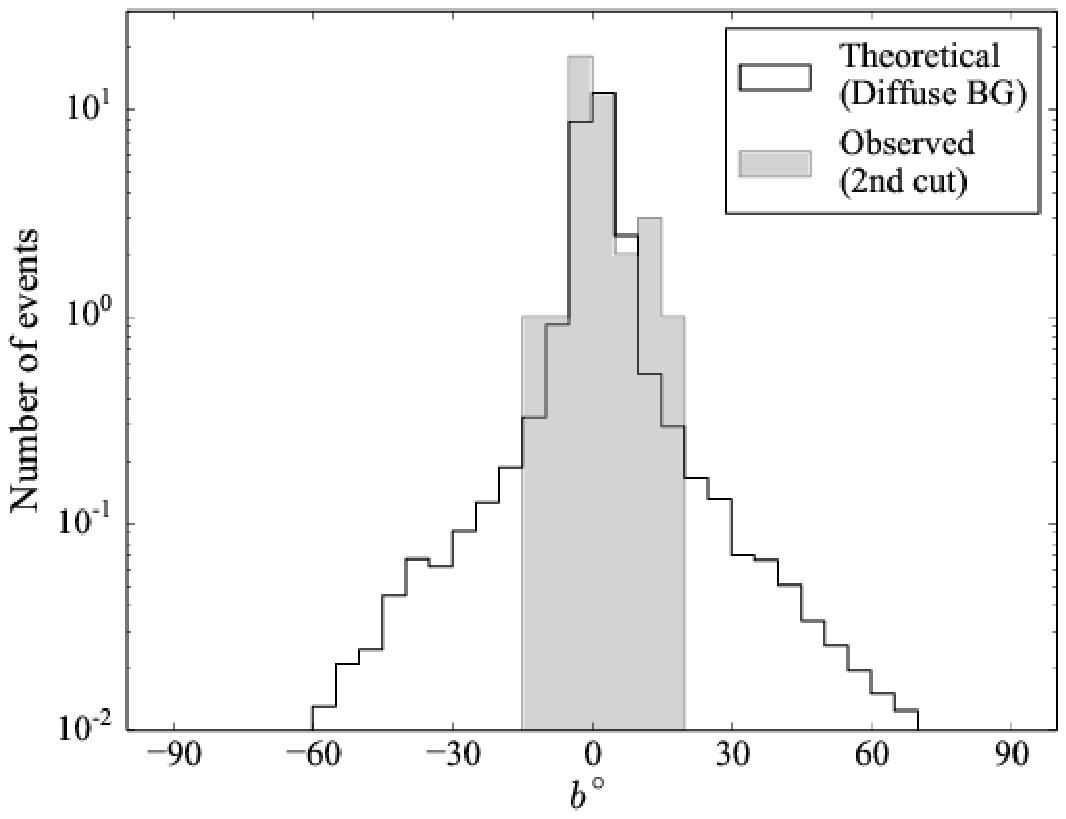}
 \caption{The Galactic latitude distribution of the MGF candidates
   with $N_{\rm{ph}}=2$ and $\Delta t = 10$ ms after the 2nd cut (gray
   histogram).  The expected distribution from the Poisson statistics
   of the diffuse gamma-ray background is shown as solid histogram.  
}
 \label{fig.1}
\end{figure}

The expected event numbers of false MGFs calculated in this way are
compared with the observed numbers after the 2nd cut in Table
\ref{table:MGFs}, and also in Fig. \ref{fig.1} as a histogram along
the Galactic latitude.  The agreement between the expected and
observed numbers is good, and hence most of the observed $N_{\rm ph} =
2$ events after the 2nd cut are considered to be produced by Poisson
statistics of the diffuse background. It should be noted that all
these events are found at low latitude regions of
$|b|<20^{\circ}$. Therefore we make the final 3rd cut to remove events
of $|b|<20^{\circ}$. Then the final result is that we found no MGF
candidates after the three event cuts, and this will be used to set a
limit on the gamma-ray/radio flux ratio of FRBs in section
\ref{sec:Implications for FRBs}.

\section{Implications for FRB\lowercase{s}}
\label{sec:Implications for FRBs}

\subsection{Modeling the population of FRBs}
\label{subsec:Modeling the population of FRBs}

To set limits on physical quantities of FRBs by the no detection of
MGFs, we first model the luminosity and rate evolution of FRBs.  The
luminosity function of FRBs is totally unknown, and here we simply use
the standard candle approximation.  Based on the typical peak flux
$F_\nu=0.5$ Jy of FRBs at observing frequency of $\nu=1.4$ GHz, and
taking $z \sim 0.5$ as FRB redshifts (\citealt{2015MNRAS.447.2852K}),
we set the radio luminosity of FRBs to be  $(\nu L_{\nu})_r =1.1\times10^{43}\rm{\,erg\:s^{-1}}$ at the rest-frame reference
frequency $\nu_r=2.1$ GHz. 

To relate gamma-ray flux and radio flux of FRBs, 
we define gamma-ray to radio luminosity ratio in the 
rest-frame of a FRB as
\begin{eqnarray}
\label{eq:xi}
\xi\equiv\frac{(\nu L_{\nu})_{\gamma}}{(\nu L_{\nu})_{r}},
\end{eqnarray}
where gamma-ray luminosity is at a gamma-ray energy of 1.5 GeV
(corresponding to the minimum Fermi gamma-ray energy 1 GeV adopted in
this work for $z = 0.5$), and radio luminosity at $\nu = \nu_r$.  We
assume that $\xi$ is the same for all FRBs, and that FRBs have a
power-law gamma-ray spectrum with a photon spectral index $\Gamma$,
i.e.,
$dL_{\gamma}/d\epsilon\propto \epsilon^{-\Gamma}$, where
$dL_{\gamma}/d\epsilon$ is the differential gamma-ray photon
luminosity per unit gamma-ray energy at the rest-frame.  Then we can
calculate the gamma-ray flux $dF_{\gamma}/d\epsilon_{\rm obs}$ as a
function of observed gamma-ray energy $\epsilon_{\rm obs}$ for a FRB
at a redshift $z$ if $\xi$ and $\Gamma$ are specified.  Since there is
no knowledge of $\Gamma$ observationally, we assume $\Gamma = 2.0$. If
we change $\Gamma$ by $\pm 2$, the final constraint on $\xi$ changes
only by about 2\%.

We model the comoving FRB rate density and its evolution as
$\Phi_{\rm{FRB}}(z,\beta)=\Phi_0 (1+z)^{\beta}$, and the 
rate in all sky for FRBs in a redshift range $z_1 < z < z_2$ is 
\begin{eqnarray}
\mathcal{R}_{\rm{FRB}}(z_1, z_2; \beta)=
\int_{z_1}^{z_2}dz \ \frac{\Phi_{\rm{FRB}}}{1+z} \
\frac{dV}{dz},
\label{eq:FRB rate}
\end{eqnarray}
where $dV/dz$ is the comoving volume element per unit redshift for all
sky, and $(1+z)^{-1}$ is the cosmological time dilation factor.  For
each assumed value of $\beta$, we determine $\Phi_0$ by the
fluence-complete rate of $\mathcal{R}_{\rm{FRB}}(0.19<z<0.89)\sim
2.5_{-1.6}^{+3.2}\times10^3\:\rm{day^{-1}}$ based on 9 FRBs (8 by
Parkes, 1 by Arecibo) at $0.19<z<0.89$ with fluence above 2
$\rm{Jy\:ms}$ \citep{2015MNRAS.447.2852K}.  The value of $\Phi_0$
changes by a factor of less than 1.5 if we use another similar rate
estimate of $\mathcal{R}_{\rm{FRB}}(0.35<z<1.3)\sim
6_{-3}^{+4}\times10^3\:\rm{day^{-1}}$ (for 9 Parkes FRBs, not
fluence-complete down to 0.9 Jy ms) by \citet{2015arXiv151107746C}.
We will present results for a range of $\beta$ = 0--4.  If FRBs are
produced by short life-time phenomena from star formation (e.g., core
collapse supernovae or young magnetars), their rate should trace the
star formation history in the universe ($\beta \sim 3$--4 to $z \sim
1$), while a smaller value ($\beta \sim 2$) is expected if FRBs are
mergers of binary neutron stars or white dwarfs
\citep{Totani:1999fk,2008PASJ...60.1327T,2013ApJ...779...72D,2014ARA&A..52..415M}.

\subsection{Upper-limit on gamma-ray / radio luminosity ratio}
\label{subsec:The upper-limit on gamma-ray / radio luminosity ratio}

Now we can calculate the expected number of MGFs by FRBs as a function
of $\xi$, thus deriving the upper bound on this parameter by the
observational result of no MGF detection for $N_{\rm ph} = 2$ MGF
events after the 3rd event cut. We have searched MGF events with four
values of $\Delta t$, but here we use $\Delta t = $ 3 msec to match
the typical observed duration of FRBs.  In a similar way to the
calculation in section \ref{subsec:BG MGFs}, we again calculate for
each pixel of the sky to a direction $\Omega_j$ in each 10-min bin
$T_i$ of observing time.  The expected number of detectable gamma-rays
from a FRB to a direction $\Omega_j$ at redshift $z$ in a MGF search
time interval $\Delta t$ is given as
\begin{eqnarray}
N_\gamma = \Delta t \int_{\rm 1 \ GeV}^{\rm 100 \ GeV} d\epsilon_{\rm obs} \ 
A_{\rm eff}(T_i, \Omega_j) \frac{dF_\gamma(\epsilon_{\rm obs};
z, \xi)}{d\epsilon_{\rm obs}} \ .
\end{eqnarray}
For a search of MGFs of $N_{\rm ph}$ photons, we can define the
maximum redshift $z_{\max}(\xi, T_i, \Omega_j)$ within which a FRB
should be detected as a MGF event, by equating $N_\gamma = N_{\rm
  ph}$.  Then the expected number of all MGF events caused by FRBs can
be calculated as
\begin{eqnarray}
N_{\rm MGF}(T_i, \Omega_j) = \mathcal{R}_{\rm FRB}[0, 
z_{\max}; \beta] \ \frac{\Delta \Omega_j}{4 \pi} 
\ T_{\rm exp}(T_i) \ ,
\end{eqnarray}
and the final total expected number of MGFs in all sky and the
observing period, $N_{\rm MGF}^{\rm tot}$, is obtained by summing up
$N_{\rm MGF}(T_i, \Omega_j)$ for $T_i$ and $\Omega_j$.  The event cuts
about the Fermi point source regions and Galactic latitude can be
taken into account in $\Delta \Omega_j$.

The expected number $N_{\rm MGF}^{\rm tot}$ increases with increasing
$\xi$ because FRBs can be detected to larger distances.  Then no MGF
detection after the 3rd event cut sets an upper limit on $\xi$, which
is estimated by equating $N_{\rm MGF}^{\rm tot}(\xi) = 3.09$,
corresponding to 95\% upper limit on the expected value of the
Poisson statistics.  The results are summarized in Table
\ref{table:xi}, and the upper limits are found to be $\xi = $
(4.2--12) $\times 10^7$ for $0 \le \beta \le 4$.  The maximum redshift
$z_{\max}$ depends on the direction to a FRB with respect to the {\it
  Fermi}-LAT FoV center, and it is $0.02$ for $\xi\sim10^8$ when a FRB
is located at the FoV center.  
This is much smaller than that of FRBs detected in radio bands ($z
\gtrsim 0.5$), confirming that {\it Fermi}-LAT is sensitive to FRBs at
smaller distances than those detected in radio bands.
Correspondingly, our limit on $\xi$ is stronger than that obtained
from non-detection of radio-detected FRBs by {\it Fermi}-LAT
(typically $\xi \lesssim 4 \times 10^{10}$).  For comparison,
\citet{2016arXiv160202188T} investigated \textit{Fermi}-GBM, 
\textit{Swift}-BAT and \textit{Konus}-WIND data at detected 16 FRB fields, setting an upper
limit on soft gamma-ray to radio flux ratio of $\xi_{\rm soft \gamma}
\lesssim 10^8$--$10^{10}$.

\begin{table}
\begin{center}
\caption{The 2$\sigma$ upper limits on gamma-ray 
to radio flux ratio ($\xi$) for $\beta$ = 0--4. 
The normalization of the volumetric FRB rate 
at $z=0$ ($\Phi_0$) is also shown.
FRB duration is assumed to be $\Delta t$ = 3 msec.
}
\label{table:xi}
\begin{tabular}{cccc}
\hline\hline
$\beta$&&$\Phi_0$ & $\xi$\\
&&$[\rm{Gpc^{-3}\:yr^{-1}}]$ 
 &$[\>\equiv\left(\nu L_{\nu})_{\gamma}\right/ (\nu L_{\nu})_{r}\>$]\\
\hline 
0&&$1.2_{-0.8}^{+1.5}\times10^4$&$<4.2_{-1.7}^{+2.4} \times10^7$\\
1&&$7.4_{-4.7}^{+9.4}\times10^3$&$<5.2_{-2.0}^{+4.3}  \times10^7$\\
2&&$4.5_{-2.9}^{+5.7}\times10^3$&$<7.0_{-2.8}^{+5.0} \times10^7$ \\
3&&$2.7_{-1.7}^{+3.5}\times10^3$&$<9.2_{-3.8}^{+7.8} \times10^7$\\
4&&$1.6_{-1.0}^{+2.1}\times10^3$&$<1.2_{-0.4}^{+1.2} \times10^8$ \\\hline
\end{tabular}
 \end{center}
\end{table}

\subsection{Discussion}

The extremely high brightness temperature of FRBs implies that their
radio emission is likely coherent radiation, and it is reasonable to
expect that it is similar to that of radio pulsars. The gamma-ray to
radio ratio is $\xi=10^4-10^8$ for gamma-ray pulsars in the Second
Fermi Large Area Telescope Catalog of Gamma-Ray Pulsars (2PC;
\citealt{2013ApJS..208...17A}). Some radio pulsars exhibit radio
efficiencies of $\sim10^{-2}-10^{-4}$ with respect to spin-down
luminosity \citep{2014ApJ...784...59S}, indicating that $\xi$ should
be smaller than $10^4$ assuming that gamma-ray luminosity cannot
exceed the spin-down luminosity. Therefore $\xi$ of pulsars are
distributed in a wide range, and our upper limit of $\xi \lesssim
10^8$ does not exclude a possibility that FRBs have a similar GeV
gamma-ray to radio flux ratio to pulsars.  Recently Repeating bursts
were discovered from FRB 121102
\citep{2014ApJ...790..101S,Spitler:2016uq,2016arXiv160308880S}, and
the top candidate progenitor model would be super-giant pulses from
pulsars. This is also consistent with our results if the gamma-ray to
radio flux ratio of giant pulses are similar to that of normal pulsar
emission.

Giant flares from soft gamma-ray repeaters (magnetars) are also
discussed as a popular hypothesis for the FRB origin
\citep{popov2010hyperflares1465953,2013Sci...341...53T,2014MNRAS.442L...9L,2014ApJ...797...70K}.
\citet{2016arXiv160202188T} placed an upper limit of $<1.1$ MJy on the
radio flux by the non-detection of a radio single pulse event from the
giant flare event of SGR 1806$-$20 on December 27, 2004, and it is
interesting to compare our limit to that of this giant magnetar flare.
Unfortunately there is no direct GeV observation of the SGR 1806$-$20
giant flare in 2004, but we can estimate a reasonable flux level as
follows.  \citealt{2007AstL...33....1F} reported 20 keV -- 10 MeV
spectrum of the giant flare using the Compton reflection from the
Moon.  Though non-thermal emission was not detected for the initial
flare, a power-law component extending to 10 MeV with a photon index
$\Gamma = 1.7$ was detected in the pulsating tail phase.  The energy
flux of the non-thermal component in 1--10 MeV is 1.9\% of the thermal
component in 20--300 keV.  If we assume the same percentage for the
initial flare, the non-thermal flux of the initial flare is estimated
to be $0.21 \ \rm erg \ cm^{-2} s^{-1}$, which is much lower than the
thermal flux at $<$ MeV and hence consistent with no
detection. Assuming that the power-law spectrum is extending to GeV
with the same photon index, we estimate $\nu F_\nu$ flux at 1 GeV to
be $0.51 \ \rm erg \ cm^{-2} s^{-1}$, and combined with the radio
upper limit, we obtain a lower limit to the GeV to radio flux ratio as
$\xi \gtrsim 10^{7.5}$.  This is marginally consistent with the upper
limit on $\xi$ derived by our work, and hence our result does not
exclude a possibility that FRBs have a similar GeV to radio flux ratio
to magnetar giant flares.

Binary neutron star mergers are also proposed as a possible candidate
for catastrophic FRB events, and in this case we may expect
associations between FRBs and short gamma-ray bursts (sGRBs)
\citep{2013PASJ...65L..12T,2014ApJ...780L..21Z}. Only one short GRB
090510 was detected in our MGF search in GeV, and note that sGRBs are
efficiently detected in soft gamma-rays with a trigger time scale much
longer than the search made here. This means that our search is not
optimized to constraining sGRBs. Furthremore, the upper limits on
$\xi$ derived in this work is valid only for sources whose event rate
is similar to that of FRBs, but it is known that the event rate of
sGRBs ($\sim10\: \rm{yr^{-1}\:Gpc^{-3}}$; \citealt{Coward01102012}) is
much smaller than that of FRBs ($\sim10^4\: \rm{yr^{-1}\:Gpc^{-3}}$).
Therefore our results do not strongly constrain the possible sGRB-FRB 
connection, and it is possible that a small fraction of FRBs
are observed as sGRBs.

\section{Conclusion}
In this paper we searched millisecond-duration gamma-ray flashes
(MGFs) in the 7-year \textit{Fermi}-LAT data, motivated by the
possible gamma-ray counterpart to FRBs. Since \textit{Fermi}-LAT is
observing wider field-of-view compared with radio telescopes finding
FRBs, FRBs at shorter distances are expected to be detected in the
\textit{Fermi}-LAT, which are out of search fields of view of radio
telescopes. Therefore we performed a blind search of
multiple gamma-ray events within a circle of
point spread function and a time interval $\Delta t = $ 1--10 msec,
and found about 100 MGF candidates. There is only one event detected
with three photons or more, and it is related to a bright
short-duration gamma-ray burst GRB 090519.  Examination of other
events indicates that all of them can be explained by the Poisson
statistics of steady gamma-ray flux from bright point sources or
diffuse gamma-ray background.  After removing regions of bright point
sources, there is no MGF event in high Galactic latitude regions of
$|b|>20^{\circ}$.

We then used this result to place an upper limit on the gamma-ray to
radio flux ratio of FRBs, $\xi\equiv\left(\nu L_{\nu})_{\gamma}\right/
(\nu L_{\nu})_{r}$.  In the calculation, we need to assume the
comoving rate density and its evolution of FRBs, and assuming a
power-law type evolution, $\Phi_{\rm{FRB}} = \Phi_0 (1+z)^{\beta}$,
we derived $\xi < (4.2$--$12) \times 10^7$ for $\beta$ = 1--4.  This
limit is stronger than those obtained by no gamma-ray detection of
known FRBs by about three orders of magnitude.  The limit is
comparable with the largest values found for pulsars. However, $\xi$
of pulsars is distributed widely in many orders of magnitudes, and we
cannot exclude that FRBs have similar values of $\xi$ to pulsars in
general. We also compared our limit with the spectral energy
distribution of the 2004 giant flare of SGR 1806$-$20. The upper bound
on the radio flux for this flare is marginally consistent with our
limit of $\xi$, if a non-thermal power-law component up to GeV exists
in the initial flare phase and its energy fraction with respect to the
thermal emission is similar to that of pulsating tail phase emission.

\section*{Acknowledgements}
% Entry for the table of contents, for this guide only

We thank the anonymous Referee for the helpful remarks and suggestions. This work was supported by JSPS KAKENHI Grant Numbers 15K05018 and
40197778.

\addcontentsline{toc}{section}{Acknowledgements}

%%%%%%%%%%%%%%%%%%%%%%%%%%%%%%%%%%%%%%%%%%%%%%%%%%



\bibliographystyle{mnras}
\bibliography{mnras} % if your bibtex file is called example.bib


%%%%%%%%%%%%%%%%%%%%%%%%%%%%%%%%%%%%%%%%%%%%%%%%%%

%%%%%%%%%%%%%%%%% APPENDICES %%%%%%%%%%%%%%%%%%%%%

\clearpage % to avoid the long table breaking up the formatting examples

% Don't change these lines
\bsp	% typesetting comment
\label{lastpage}
\end{document}